\documentclass{amia}
\usepackage{graphicx}
\usepackage{booktabs}	
\usepackage{graphics}
\usepackage{graphicx}
\usepackage{adjustbox}
\usepackage{multirow}
\usepackage{amsmath}
\usepackage{bbm}
\usepackage{subfigure}
\usepackage{enumitem}

\usepackage[numbers,sort,super]{natbib}
\usepackage{color}
\usepackage[labelfont=bf,labelsep=period]{caption}

\begin{document}

\title{Visualizing Patient Timelines in the Intensive Care Unit}

\author{Dina Levy-Lambert, S.B.,$^{1}$ Jen J. Gong, Ph.D.,$^{1}$ Tristan Naumann, Ph.D.,$^{1}$ \\ Tom J. Pollard, Ph.D.,$^{1}$ John V. Guttag, Ph.D.$^{1}$}

\institutes{
    $^1$Massachusetts Institute of Technology, Cambridge, MA, USA\\
}

\maketitle

\noindent{\bf Abstract}

\textit{Electronic Health Records (EHRs) contain a large volume of heterogeneous patient data, which are useful at the point of care and for retrospective research. These data are typically stored in relational databases. Gaining an integrated view of these data for a single patient typically requires complex SQL queries joining multiple tables. In this work, we present a visualization tool that integrates heterogeneous health care data (e.g., clinical notes, laboratory test values, vital signs) into a single timeline. We train risk models offline and dynamically generate and present their predictions alongside patient data. Our visualization is designed to enable users to understand the heterogeneous temporal data quickly and comprehensively, and to place the output of analytic models in the context of the underlying data.}

\section*{Introduction}

Electronic Health Records (EHRs) contain increasingly detailed patient information and are useful both at point of care and for retrospective analysis. Underlying EHRs is typically a relational database, comprising data collected from across multiple monitoring systems. Integrating data from a relational database to obtain a patient-centric perspective requires SQL queries joining multiple tables, and understanding the relationships across these data tables can be challenging. EHRs also contain many different types of data, including unstructured clinical notes, time spent in different locations in the hospital, and time-series describing patient vital signs and lab test results. Effective integration of heterogeneous data, while challenging, would enable a comprehensive view of a patient stay. Both clinicians and researchers can benefit from tools that assist in developing a quick, accurate, and comprehensive summary of patients' data. Visualizations are a well-recognized method for summarizing patient data into useful abstractions and highlighting important information to reduce information overload.~\cite{gotz2016data}

In this paper, we demonstrate a visualization approach that integrates heterogeneous data from a patient's stay in the hospital into a single timeline. Our visualization presents vital signs, laboratory test results, unstructured clinical notes, and duration-based interventions and stays in a single picture. This allows users to explore patient data by selecting a particular patient of interest, or by applying filters to select a cohort of interest. Additionally, users are able to import a risk model in order to visualize predictions alongside the clinical data. Visualizing the underlying data alongside the evolving risk profile provides context for understanding why patient risk is changing.

When a user selects a patient, the model estimates are evaluated dynamically and populate the risk timeline. We demonstrate how integrating risk predictions from a predictive model with the associated physiological timeline can be useful in drawing visual correlations with the underlying data. Such a visualization could be applicable at the point of care, when a clinician wants to quickly understand the patient status from notes by other care staff alongside the patient's vital signs and intervention data. It can also be useful for researchers who are exploring a clinical database, or who, for example, want to understand a model's risk predictions in the context of the underlying data.

\begin{figure}[t!]
\centering
\includegraphics[scale=0.27]{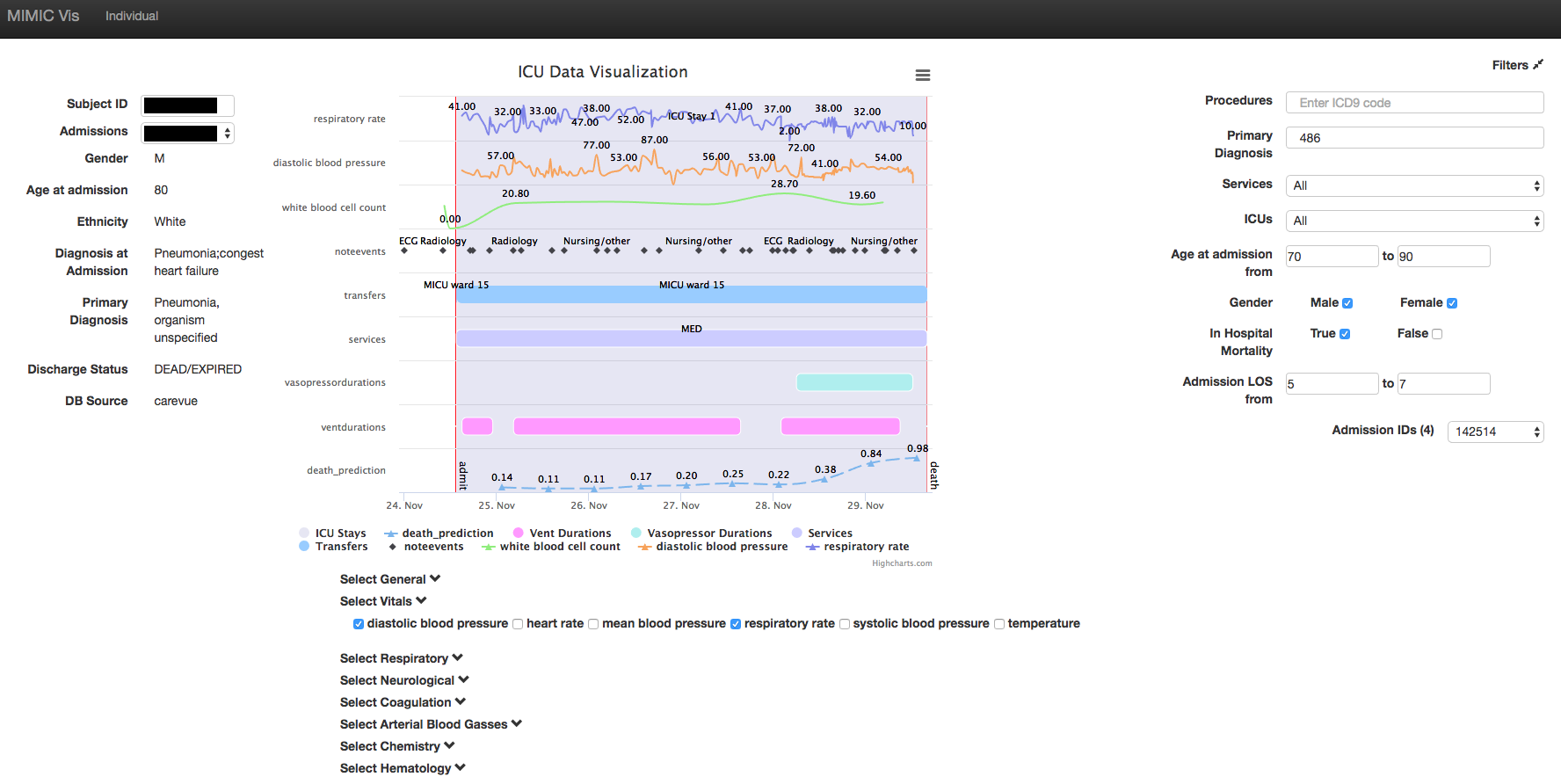}
\caption{A visualization generated from the MIMIC-III critical care dataset. A patient with diagnoses of pneumonia and congestive heart failure is shown with several vital signs and laboratory test values, alongside a risk profile. Clinical notes are displayed in a time-series of points indicating note category, and the corresponding text appears when the user hovers over the point. The time of admission and time of death are marked with vertical red lines. Static information about the admission is available to the left of the timeline, filters to identify a cohort of interest are on the right, and checkboxes for time-series of interest are below. Tooltips are revealed when the user hovers over a particular point, indicating additional information about the values shown.
Per the MIMIC-III data use agreement, the specific subject and admission are redacted, and text from clinical notes is not shown.}
\label{full_demo}
\end{figure}

We demonstrate our visualization approach through a web-based tool we built for the Medical Information Mart for Intensive Care dataset (MIMIC-III).~\cite{johnson2016mimic} MIMIC-III is a publicly available dataset that contains data extracted from Beth Israel Deaconess Medical Center's EHR over the years 2001--2012. While our implementation is built for use with MIMIC-III, we believe this approach of visualizing heterogeneous data and integrating output from machine learning algorithms is generally useful for EHR data. A visualization generated from MIMIC-III is shown in Figure~\ref{full_demo}. 

Our contributions are as follows:
\begin{enumerate}[nolistsep]
\item We present a visualization approach centered around hospital admissions that is generally applicable to EHR data. In particular, our approach is designed to integrate heterogeneous data into one flexible visualization.
\item We present a web-based visualization tool for patient data in the publicly available MIMIC-III database. Our visualization demonstrates how intervention data, hospital location data, physiological time-series, and clinical notes can be viewed on a single timeline. Our tool will be made publicly available.
\item We develop and evaluate predictive models for in-hospital mortality at different points during a patient's ICU stay. Models are applied to a selected patient's data dynamically. The risks are then integrated into the patient timeline. We demonstrate how this tool can be used to relate changes in risk with observations from the stay.

\end{enumerate}

In the following sections, we present related work, a description of the visualization approach, specifics about the tool, an overview of the predictive model development and evaluation, and a case study demonstrating the utility of our tool in understanding patient stays.

\section*{Related Work}
\label{related_work}


Existing clinical visualization tools generally focus either on identifying and summarizing patient cohorts,~\cite{stubbs2012sim,malik2015cohort,lee2016web,harris2016i2b2t2,wongsuphasawat2011outflow} or on visualizing patient timelines.~\cite{lifelines,lifelines2,gotz2014methodology} Visualization tools that focus on a temporal understanding of patient data use different types of data, including event sequences that ignore values (e.g., lab test performed rather than lab test results),~\cite{lifelines, lifelines2, gotz2014methodology} physiological time-series (e.g., heart rate and blood pressure over time),~\cite{stubbs2012sim} and clinical notes.~\cite{hirsch2014harvest} However, these works mostly focus on a {\it single} type of data, rather than integrating heterogeneous data. In contrast, our visualization integrates clinical notes alongside physiological time-series and duration data describing treatments and stays.


Few works have used predictive models to augment the existing visualization. \citeauthor{gotz2014methodology} mined event sequences containing diagnosis codes, medications, procedures, and lab tests, and identified commonly occurring patterns that are correlated with a user-specified outcome.~\cite{gotz2014methodology} These sequences, which do not make use of associated values (e.g., lab test results) are then highlighted in their visualization. In contrast to this work, we use values-based data such as physiological time-series and we demonstrate how machine learning models trained and evaluated offline can be used to dynamically generate an additional picture of risk over time for a selected patient.




Our work focuses on the MIMIC-III dataset. While few visualizations have been proposed for this dataset, several exist for previous versions.
\citeauthor{chen2015explicu} present a web-based visualization tool for MIMIC-II.~\cite{chen2015explicu} Their visualization incorporates both a patient timeline module as well as a predictive modeling module. Like our work, the authors demonstrate how a visualization of a predictive model for mortality can be used for additional insight. Unlike our work, however, the authors investigate a single static prediction task, rather than looking at how estimates of risk for a given patient vary over the stay. Additionally, they display a smaller subset of the available data in MIMIC-II (diagnosis codes, procedures, and medications) compared to our visualization. Lastly, their proposed visualization isolates predictive modeling results from the patient data; instead, we integrated these results into the timeline.

\citeauthor{lee2016web} also provide a web-based visualization tool for MIMIC-II.~\cite{lee2016web} However, their visualization focuses on aggregate statistics in order to identify cohorts of interest. The authors propose this tool as a useful preliminary research tool for users who are new to MIMIC. However, their tool does not enable the researcher to understand the specific attributes of a patient stay that are unique to MIMIC. 

\subsection*{Visualization}
\label{viz}


Our visualization consists of four panels: 1) a patient filtering panel (Figure~\ref{full_demo}, right), 2) a patient attribute panel, 3) a time-series selection panel (Figure~\ref{full_demo}, bottom), and 4) the patient timeline panel (Figure~\ref{full_demo}, center).

A typical workflow starts by identifying admissions satisfying certain criteria using the filtering panel, and selecting one of these to visualize. After a visit is selected, the default time panel appears showing intervention durations, horizontal bars depicting time spent in different areas in the hospital (Figure~\ref{full_demo}, left), indications that a clinical note has been entered, and the results of the risk prediction models described in the next section of this paper. The next step is checking those boxes in the time-series selection module that are of interest. We now discuss each of these steps in more detail.



\begin{figure}[b!]
\centering
\fbox{\includegraphics[scale=0.3]{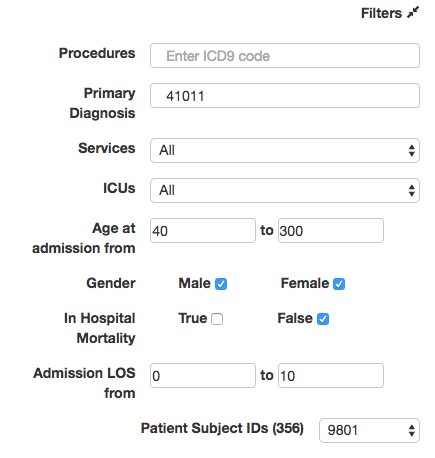}}
\hspace{1.5cm}
\fbox{\includegraphics[scale=0.3]{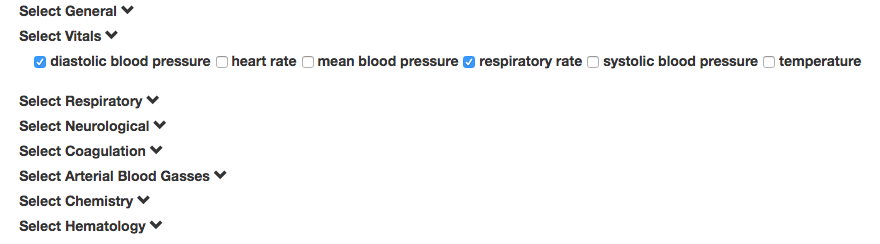}} 
\caption{Filtering panel to identify patient admissions with particular characteristics ({\bf A}), and checkboxes for indicating which time-series to display on the timeline ({\bf B}). Checkbox categories and the filters panel can be minimized to allow the user to focus on the data displayed on the timeline after an admission has been chosen.}
\label{filters}
\end{figure}

\textit{Patient Filtering Panel:}
Patient selection can be done by either specifying a subject identifier (see Patient Attribute Panel), or by indicating criteria upon which to filter.  Filters facilitate the selection of patients when specific subject identifiers are not known. We provide several commonly used patient filters, including procedures and primary diagnoses (using ICD-9 codes), services received, age at admission, gender, length of stay, and mortality (Figure~\ref{filters}A). A drop-down menu in the patient attribute panel enables a user to investigate other admissions corresponding to the selected patient. Once a particular admission has been selected, the filtering module can be minimized.

\textit{Patient Attribute Panel:}
This panel contains static attributes about a patient (e.g., gender, age, admitting diagnosis). Subject identifiers can be entered into the appropriate field. If a subject identifier is directly entered, the timeline by default shows the patient's first admission. All other fields in this panel are read-only and are automatically filled in.

\textit{Patient Timeline Panel:}
A single patient may have upwards of thousands of data points stored across the various tables for a single admission. Our visualization combines data from many tables, and chooses distinct ways to display data of differing types, including lab values and vital signs over time, interventions, and unstructured clinical notes. Single events in time (e.g., time of admission, time of discharge) are visualized with vertical red lines. Duration-based data such as location in the hospital or interventions are shaded or shown as horizontal bars indicating time from start to end. Clinical notes are shown as points corresponding to the date or time of entry, and the text is revealed when the user hovers over the point. Time-series for lab test values, other physiological measures, and risk assessments are shown as solid lines.

Tooltips for time-series data also appear when the user hovers over specific points on the time-series (e.g., type of fluid a lab test was performed on). These design decisions highlight the differences between the different types of available data, and reveal the richness of the data without overwhelming the user. 
Each time-series is shown on a separate vertical axis for ease of readability. The axis of the timeline expands to accommodate additional attributes selected by the user. Users can focus on specific times and attributes by selecting regions to zoom in on.

\textit{Time-series Selection Panel:}
The set of time-series that are displayed can be selected using checkboxes (Figure~\ref{filters}B). The time-series are organized by category (e.g., blood gasses) that can be minimized or expanded. Most time-series appear in only one category, but some (e.g., oxygen saturation) appear in multiple categories.


\textit{Visualizing Risk Over Time:}
Often, researchers are interested in understanding a patient's risk for adverse events.
This risk can take the form of established risk scores (e.g., SAPS-II), or the predictions of a novel method under development. In both cases, risk scores and their evolution over a patient stay can give users additional understanding of patient state. We present changing patient risk as an additional time-series on the same plot as the other time-series attributes. This integration with the existing heterogeneous data allows users to draw visual correlations with the underlying data.

To our knowledge, no visualization has integrated risk timelines with the existing patient data in a patient timeline. Our approach is patient-centric, to emphasize what care is being provided to the patient, how a patient's physiology changes over time, and how these factors accumulate into changing risk of adverse outcomes.


In a later section of the paper, we present a case study illustrating how these pictures of patient stays can be used. 

\textit{MIMIC-III:}
We demonstrate our visualization on MIMIC-III. In this database, vital signs are only available while a patient is in the ICU, but lab tests and clinical notes are available during the entire hospital admission. ICU stays are identified by unique IDs, which are associated with corresponding \textit{admission IDs} and patient \textit{subject IDs}. These IDs are used to link information about a patient's stay across otherwise isolated tables.

Our timeline captures several examples of each type of data previously mentioned (e.g., major lab tests, vital signs, all clinical notes, and selected interventions), and the approach is easily generalizable to new examples of these categories of data.

\subsubsection*{Construction and Validation of Risk Models}
\label{risk_models}

In the previous section, we described how risk models, learned offline, can be imported to dynamically generate a specific patient's risk timeline and visualized alongside the existing heterogeneous data. In this section, we describe an example of such a risk model that we constructed and validated offline.

We used machine learning to construct models to predict in-hospital mortality for patients over the age of 15. In keeping with our philosophy of viewing a hospital stay as a timeline, we trained and evaluated a different risk model at every twelve hours for the first seven days of the first ICU stay, resulting in a total of 14 models. This framework is presented in \citeauthor{ghassemi2014unfolding}~\cite{ghassemi2014unfolding} Our models predict whether a patient will die at least twelve hours after the time the prediction is made. This is done so that if the models were used prospectively there would be time to intervene.

\begin{figure}[t!]
\centering
\includegraphics[scale=0.35]{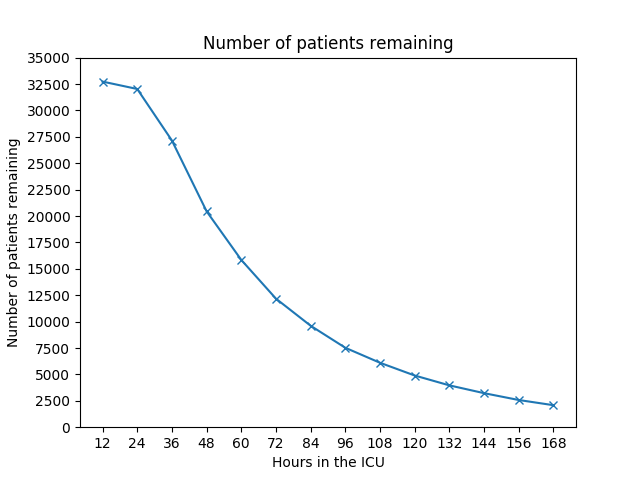}\includegraphics[scale=0.35]{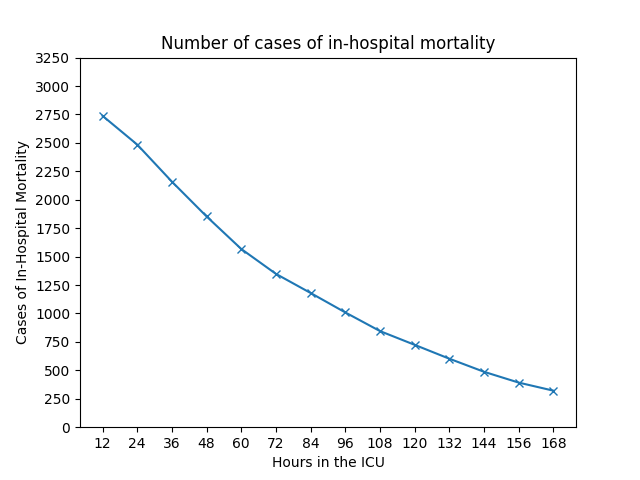}\\ 
\includegraphics[scale=0.35]{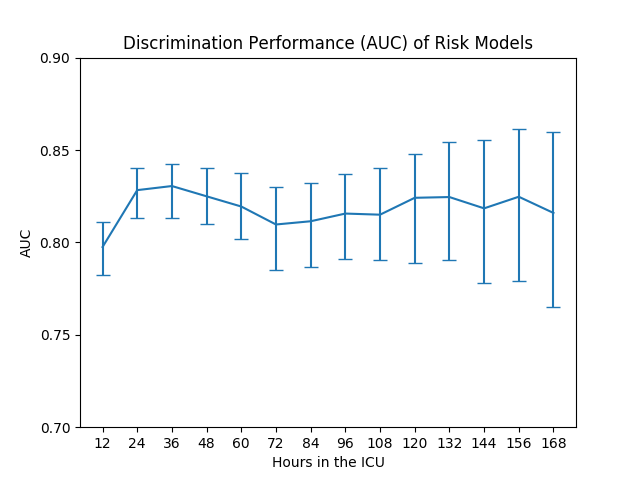}\includegraphics[scale=0.35]{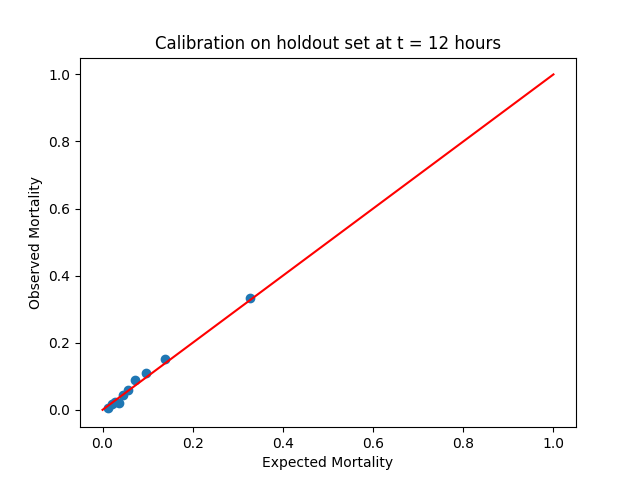}
\caption{Number of patients (top left, \textbf{A}) and number of adverse events (top right, \textbf{B}) over time. Discrimination performance (AUC, bottom left, \textbf{C}) and calibration (bottom right, \textbf{D}) of mortality prediction models for prediction tasks using 12 hours to 168 hours of information from patient ICU stays in increments of 12 hours, with a gap of 12 hours prior to outcome.}
\label{results}
\end{figure}

The models were built using L2-regularized logistic regression and 29 physiological features (vital signs and lab test values) from each 12 hour window. Since the amount of available information grows over the course of a stay, the number of features in each model grew. Thus, at 12 hours into the patients' ICU stays, the model had 29 features, and at 24 hours into the patient stay, the model had 58 features. The features we used are listed in Table~\ref{features}. We implemented these models starting at 12 hours into the patients' stays, and up until 7 days (168 hours) to maintain an adequate support for the model (shown in Figures~\ref{results}A~and~\ref{results}B), resulting in a total of 14 models.  The regularization parameter was learned using 5-fold stratified cross-validation on the training set, and an asymmetric cost parameter was used to combat the class imbalance during learning. All models were implemented using scikit-learn.~\cite{scikit-learn} To generate a risk timeline for the visualization, the models are applied to the admission of interest and the risk estimates are joined by a dashed line to indicate a time-series.

\begin{table}[t!]
\centering
\caption{Features included in the risk models.}
\begin{tabular}{|c | c |} \toprule
\multicolumn{2}{| c |}{\bf{Features}} \\ \hline
{\bf Vitals} & diastolic blood pressure, systolic blood pressure, mean blood pressure, \\
& Glascow Coma Scale total, heart rate, respiratory rate, temperature, \\
& weight, white blood cell count, pH \\ \hline
{\bf Labs} & anion gap, bicarbonate, blood urea nitrogen, chloride, \\
& creatinine, fraction inspired oxygen, glucose, hematocrit, \\
& hemoglobin, INR, lactate, magnesium, oxygen saturation, \\
& partial thromboplastin time, phosphate, platelets, \\
& potassium, prothrombin time, sodium \\ \bottomrule
\end{tabular}
\label{features}
\end{table}

Although logistic regression typically returns well-calibrated
probabilities, our use of an asymmetric cost parameter to weight the rarer, adverse events as equally important to the larger class of non-adverse events creates a bias in the model. This is because the model no longer sees the true incidence of events in the population. Because of this, after model training, we used Platt scaling to calibrate each model's probability estimates.~\cite{platt1999probabilistic} Calibrated models are important in the context of individual patient risk prediction,
especially because we are integrating the risk estimates from multiple different models into a single patient timeline. We evaluated our model in terms of both discrimination (Area Under the Receiver Operating Characteristic curve (AUC)) and calibration (Hosmer-Lemeshow test~\cite{hosmer2013applied}). These results are shown in Figure~\ref{results}.

Our model demonstrates consistent AUC ($\approx 0.8$) across all
prediction tasks (Figure~\ref{results}C). The confidence intervals become wider over time, because of the decreasing support for the model as patients who are discharged or die during the time interval of interest are removed from model training and testing. The model support is shown in Figures~\ref{results}A and ~\ref{results}B. Our models also demonstrate reasonable calibration; the calibration of the model trained using 12 hours of data is shown in Figure~\ref{results}D. The points in the calibration plot indicate the observed incidence of mortality in the deciles of the predicted risk. This plot appears to show that expected and observed mortality lie along the 45 degree line, indicating good calibration. We additionally computed the $p$-values from the Hosmer-Lemeshow test to evaluate goodness of fit. Of the 14 models considered, 4 had a $p$-value below 0.01, and 4 others had a $p$-value below 0.05. The less well calibrated models used less information
(e.g., t = 12, 24, 36 hours) or had a lower support (e.g., t = 168 hours). 

Though our performance is competitive with other published models, our goal was not to produce the best possible model.  Other more complex models and features could be used to achieve better discriminative performance (e.g., \citeauthor{ghassemi2014unfolding},~\cite{ghassemi2014unfolding,ghassemi2015multivariate}, \citeauthor{che2016recurrent},~\cite{che2016recurrent}~\citeauthor{wiens2012learning}~\cite{wiens2012learning}). Our
goal was to demonstrate how integrating individual patient risk in the timeline can be used to understand the risk model in the context of the clinical data. Importantly, our tool allows users to import their own model or specific patient risk scores.

\section*{Case Study}
\label{case_studies}

\begin{figure}[t!]
\centering
\includegraphics[scale=0.2]{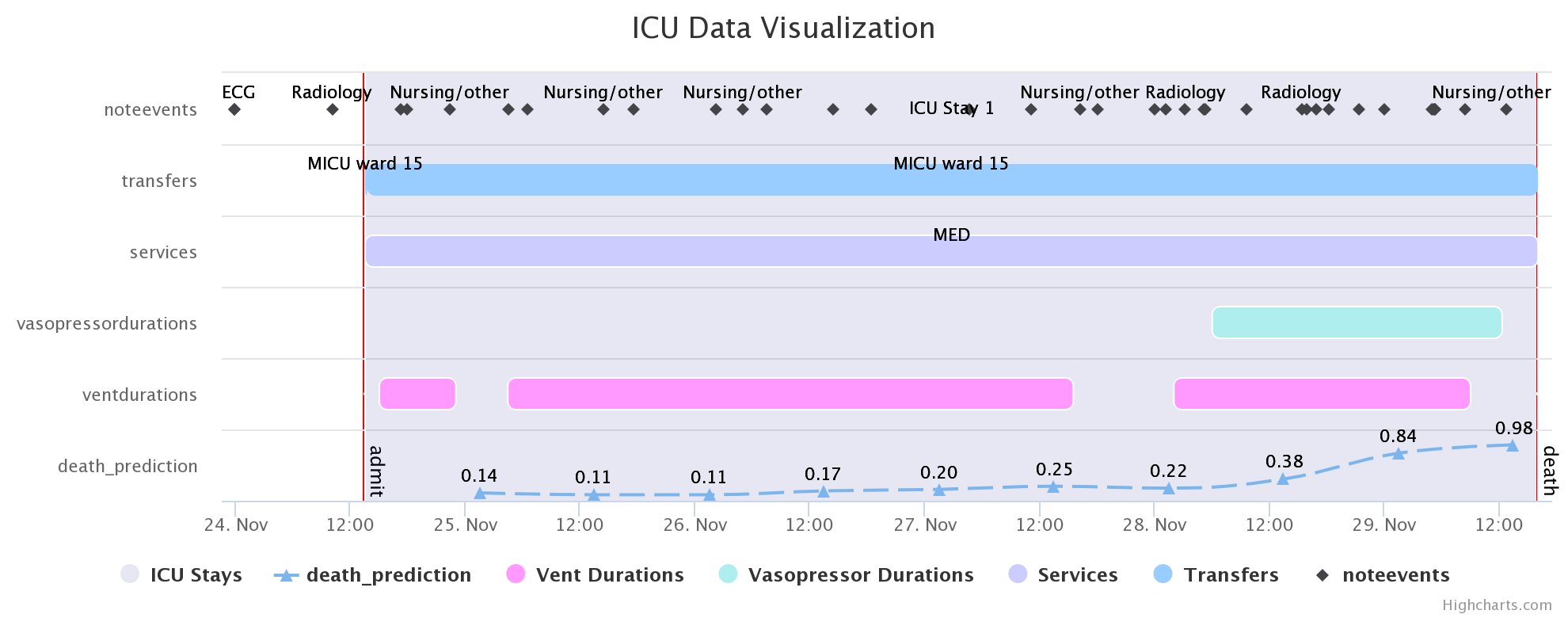}
\caption{The initial timeline for an 80-year-old male, admitted with pneumonia and congestive heart failure. The patient shows steadily increasing risk during the course of his stay until his death.}
\label{chart_0}
\end{figure}

In this section, we present a case study demonstrating the use of our visualization tool. We focus on examining a single patient, an 80-year-old man who was admitted to the hospital with diagnoses of pneumonia and congestive heart failure. This patient died in the hospital, approximately six days into the stay. When the patient is selected, his default patient timeline appears (Figure~\ref{chart_0}). This timeline tells us that the patient was put on a ventilator shortly after admission, and that he stayed in the Medical ICU for the duration of the admission. The patient was given an echocardiogram, as indicated by the echo report in his clinical notes time-series. Initially, the patient's estimated risk of mortality is low. However, over the course of his stay, and in particular during the day of November 28, his risk increases dramatically. This follows a brief period of time during which, as the visualization shows, the patient was taken off the ventilator. Shortly after the risk starts to rise, the visualization indicates the initiation of vasopressors.

\begin{figure}
\centering
\includegraphics[scale=0.2]{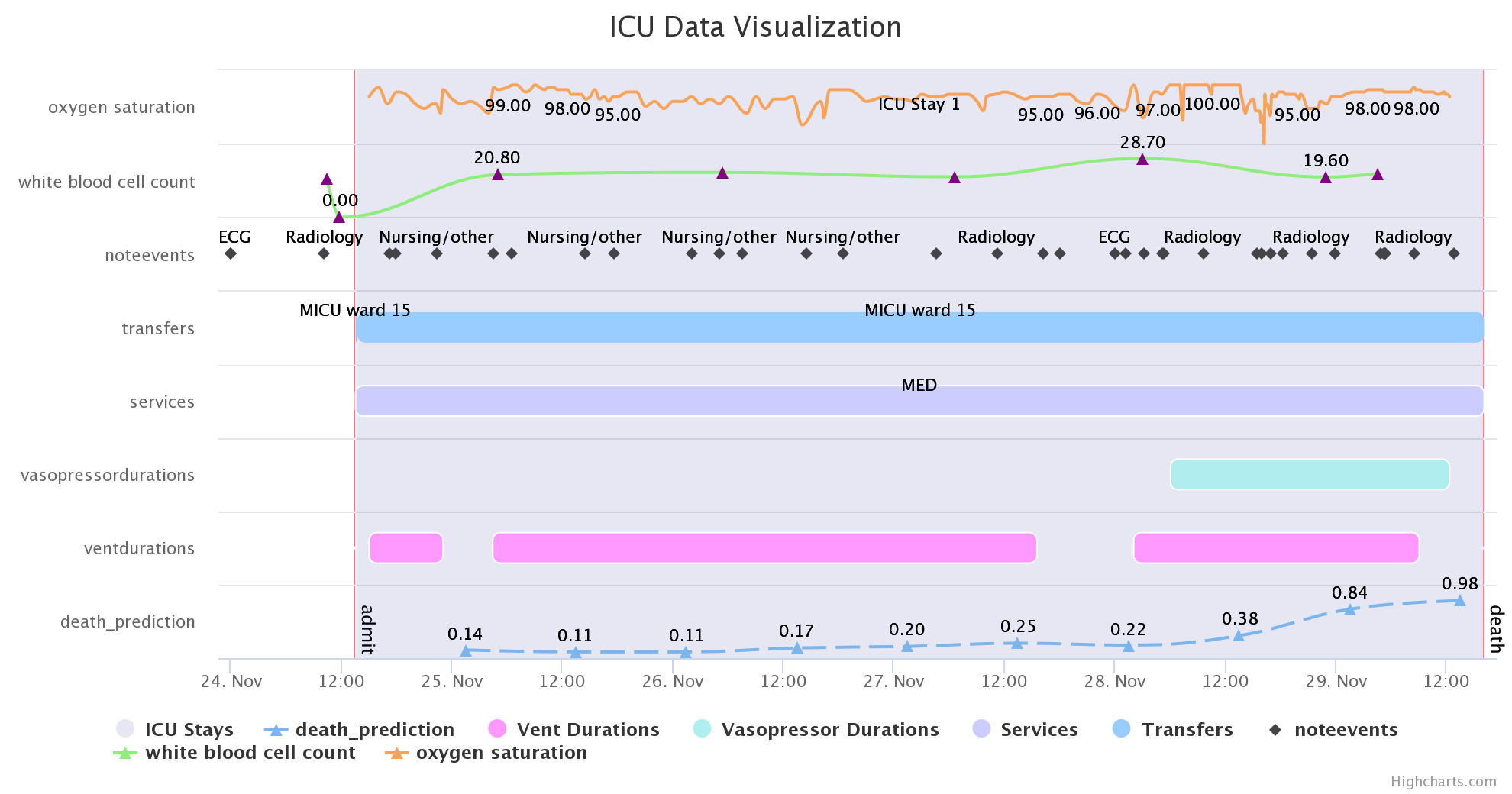}
\caption{The patient timeline expands after adding oxygen saturation and white blood cell (WBC) count. The patient shows elevated WBC count and drops in oxygen saturation when the patient is taken off of the ventilator.}
\label{chart_1}
\end{figure}

The next step is to use the time-series selection panel to examine potentially relevant variables. Based on the diagnoses we decided to start by visualizing white blood cell (WBC) count and oxygen saturation (Figure~\ref{chart_1}). The timeline axis expands to accommodate the additional time-series, so that the signals are not crowded. From observing these time series, the patient appears to have an anomalous WBC count prior to admission, but the first record while in the ICU is elevated. This agrees with his admitting diagnosis of pneumonia. In addition, we note that the oxygen saturation stays relatively level while the patient is on the ventilator. Sometime after ventilation was stopped, the patient's  oxygen saturation dipped to 86\%, after which he was put back on the ventilator.

\begin{figure}
\centering
\includegraphics[scale=0.2]{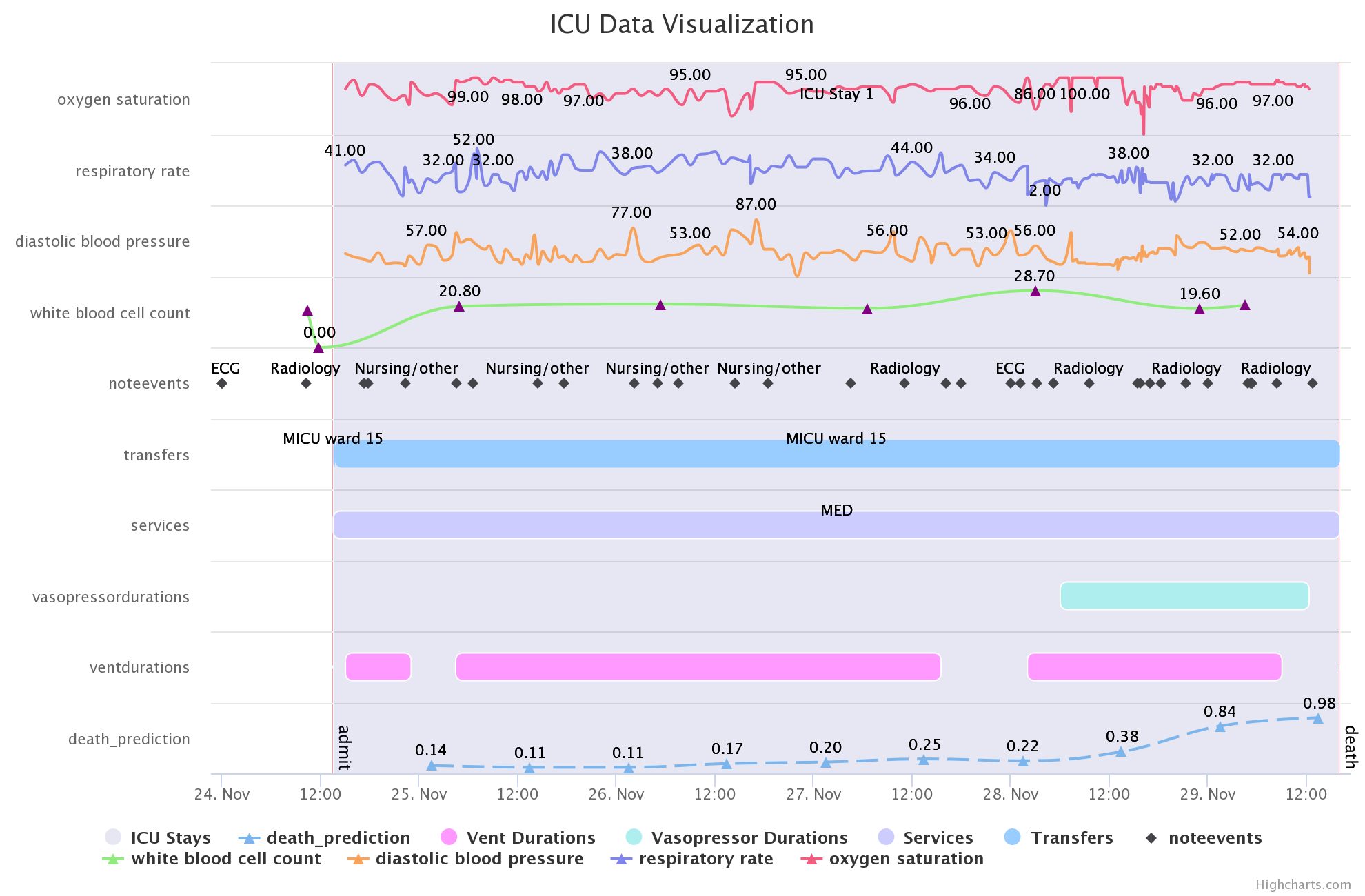}
\caption{Respiratory rate and diastolic blood pressure are added to the patient timeline. A drop in the respiratory rate occurs around the time the patient is taken off of the ventilator, and the onset of vasopressors corresponds to an increase in diastolic blood pressure.}
\label{chart_3}
\end{figure}
We next tried to understand what was happening around the time that the patient's risk started to climb. We observed that shortly after the risk started to rise, the WBC count increases to 28.7, the highest during the admission. We also observed that the oxygen saturation seemed erratic. Next, we checked the diastolic blood pressure and respiratory rate (Figure~\ref{chart_3}) boxes in the time-series selection panel. Blood pressure was of interest because the patient was put on vasopressors during this time, indicating a low blood pressure. In addition, we were interested in observing how respiratory rate was affected by being taken off of the ventilator for a brief period. Our risk model made use of physiological measures in successive 12 hour periods, and the underlying patient data correlate with this increase in risk. We used the tooltips feature to look more closely at the respiratory rate over this period, and noted that it dips to 2 BPM when the risk increases from 0.22 to 0.38. A clinical note at the end of the stay indicated that a decision had been made to take comfort measures only. This correlated with the patient being taken off of vasopressors and the ventilator, and with a further increase in the estimated risk. The patient's time of death was recorded shortly after.

This case study indicates how features relevant to the patient's condition can be progressively selected by the user. It also highlights how the visualization is designed to accommodate this additional information without overwhelming the user. Finally, the integration of the risk timeline with the physiological time-series and stay events allows the user to draw visual correlations to better understand what data are driving the risk model predictions.

\section*{Discussion and Conclusions}

We propose a visualization approach that integrates heterogeneous data in a patient timeline. This approach is applicable to a generic EHR dataset; we apply the method to MIMIC-III, an intensive care unit dataset that is publicly available. We integrate the output of machine learning models for predicting patient risk into these timelines. This enables users to visualize how risks relate to treatments, physiological time-series, and clinical notes during a patient's stay. We demonstrate our visualization on a case study in MIMIC-III using a simple prediction model framework. 
Our case study illustrates how a user can understand a patient's stay in the hospital through the integration of the data on a single timeline. This stands in contrast to the typical process of understanding heterogeneous EHR data stored in a relational database, where users must join multiple different tables on a given ID to obtain these data, and even then, it is difficult to visualize them in the context of a patient stay.

Our integration of in-hospital mortality risk estimates is just one illustration of how machine learning can be utilized to augment visualizations of patient data. For example, topic modeling of the clinical notes could be used to create more concise representations of the notes in the timeline. Anomalous values could be identified using machine learning or statistics and highlighted within the visualization. Additionally, we could integrate a time-series reflecting changes in the most predictive features over time.

In summary, our proposed visualization integrates heterogeneous clinical data and predictive models to augment a user's understanding of a given patient's stay. It can be used to understand data in a relational database from a patient-centric perspective. This could prove useful in a clinical setting, to help clinicians monitor patients and access heterogeneous data from a single interface, and to assist researchers in better understanding the available data and how risk models relate to the underlying data.

\section*{Acknowledgements}
The authors would like to thank Alistair Johnson from the MIT Laboratory for Computational Physiology for his input and suggestions.

This research was funded in part by the SuperUROP program at MIT, Intel Science and Technology Center for Big Data, the National Library of Medicine Biomedical Informatics Research Training grant 2T15 LM007092-22, NIH National Institute of Biomedical Imaging and Bioengineering (NIBIB) grant 5-R01-EB017205-05, and Quanta Computer, Inc.

\makeatletter
\renewcommand{\@biblabel}[1]{\hfill #1.}
\makeatother

\newpage
\bibliographystyle{vancouver-authoryear}
\bibliography{viz}



\end{document}